# Stool Recognition for Colorectal Cancer Detection through Deep Learning

Glenda Hui En Tan[1*], Goh Xin Ru Karin[2*] and Shen Bingquan[3]

[1]Carnegie Mellon University
glendat@andrew.cmu.edu
[2]London School of Economics and Political Science
22ygohx918d@alumni.ri.edu.sg
[3]DSO National Laboratories Singapore
SBingqua@dso.org.sg

## Abstract

Colorectal cancer is the most common cancer in Singapore and the third most common cancer worldwide. Blood in a person's stool is a symptom of this disease, and it is usually detected by the faecal occult blood test (FOBT). However, the FOBT presents several limitations: the collection process for the stool samples is tedious and unpleasant, the waiting period for results is around two weeks and costs are involved. In this research, we propose a simple-to-use, fast and cost-free alternative – a stool recognition neural network that determines if there is blood in one's stool (which indicates a possible risk of colorectal cancer) from an image of it. As this is a new classification task, there was limited data available, hindering classifier performance. Hence, various generative adversarial networks (GANs) (DiffAugment StyleGAN2, DCGAN, Conditional GAN) were trained to generate images of high fidelity to supplement the dataset. Subsequently, images generated by the GAN with the most realistic images (DiffAugment StyleGAN2) were concatenated to the classifier's training batch on-the-fly, improving accuracy to 94%. This model was then deployed to a mobile app – Poolice, where users can take a photo of their stool and obtain instantaneous results if there is blood in their stool, prompting those who do to seek medical advice. As "early detection saves lives", we hope our app built on our stool recognition neural network can help people detect colorectal cancer earlier, so they can seek treatment and have higher chances of survival.

# 1 Background and Purpose of Research Area

## 1.1 About Colorectal Cancer

Colorectal cancer is the most common cancer and one of the top killer cancers in Singapore. According to the Singapore Cancer Registry, about 1500 patients are diagnosed with colorectal cancer annually and around 650 of them die from it each year [1]. Worldwide, colorectal cancer is the third most common cancer and there were over 1.8 million new cases in 2018 [2].

Despite being a deadly disease, the earlier colorectal cancer is detected, the more likely it can be successfully treated. Colorectal cancer can be detected through a number of screening tests. One such test is colonoscopy, to observe for the presence of polyps (abnormal tissue growths which may turn cancerous) [3]. Another test is the faecal occult blood test (FOBT), a laboratory test that checks stool samples for blood (a common symptom of colorectal cancer). The Singapore Cancer Society provides this test [4, 5]. However, this test has several limitations. It is tedious and unpleasant (one has to collect two stool samples); the waiting period for results takes around two weeks and it is only free for Singaporeans and permanent residents aged 50 and above.

In this paper, our contributions include:
  a. Producing a Generative Adversarial Network (GAN), more specifically, a Diffaugment StyleGAN2, that produces realistic synthetic images of normal stool, stool with blood and empty toilet bowl images, to supplement the insufficient images available.
  b. Developing a stool recognition neural network which classifies stool with blood from normal stool with a high accuracy of 94%.
  c. Deploying the model to an Android mobile app - Poolice, where users take a picture of their stool and obtain instantaneous diagnosis on their risk of colorectal cancer, prompting those at risk to seek medical attention early. This app is a simple-to-use, convenient, fast and cost-free alternative to the FOBT.

## 1.2 Addressing the Insufficient Dataset

As this is a new classification task and there are currently no suitable datasets, we compiled our own dataset of 1610 images consisting of three classes - 337 stool with blood, 1038 normal stool and 235 no stool (empty toilet bowl) images. The images were sourced from online platforms such as Google and Reddit, as well as from friends and family members. However, since a large amount of training data is essential in making deep learning models successful, this lack of data may hinder our classifier's performance by increasing the probability of overfitting.

One way to increase the dataset size is image augmentation, which artificially creates new training images from existing ones through a combination of processing techniques (random rotation, height and width shifts, horizontal and vertical flips etc.) However, as the synthetic images are produced from existing images, they still share similar features. It would be more ideal for the classification model to be exposed to a wider range of data through different images. Instead, image augmentation can be used to improve on a dataset with sufficient content.

With the recent successes of GANs in generating realistic images, GANs were used to generate normal stool, abnormal stool and no stool images.

### 1.3 Generative Adversarial Networks (GANs)

Described as "the most interesting idea in the last ten years in machine learning" by machine learning expert Yann LeCun, GANs consist of a generator which generates fake images similar to the training input from random noise (usually Gaussian distribution) and a discriminator, which differentiates fake images produced by the generator from real images. During training, the generator tries to outsmart the discriminator by generating increasingly convincing fakes, while the discriminator learns to better identify the real data from the fakes. Nash equilibrium is reached when the generator generates perfect fakes, and the discriminator is left to guess at 50% confidence if the generator output is real or fake. Apart from creating DeepFakes and performing super-resolution tasks, GANs can also be used to increase the sizes of datasets, improving the performance of classifiers.

## 2 Hypothesis of Research

Our research question is: Can the use of GANs improve the performance of a classifier by increasing the size of the dataset? We hypothesize that after learning the features of the given images, GANs can generate realistic and high-fidelity images, giving rise to a larger dataset. With a larger training set, the use of GANs can improve the performance of classifier models, enabling them to achieve a higher accuracy.

## 3 Methodology

Inputs and Model Architecture: In order to ensure a fair experiment, all classification models were evaluated on a fixed test set of 160 images (60 abnormal stool, 60 normal stool and 40 no stool images). All models had the same architecture (see Appendix). All images were resized to 128x128x3 inputs and normalized to pixel range [0, 255] for the classification models.

Training Details: All classification models were trained for 200 epochs using 5-fold cross validation, a resampling procedure used for model evaluation on a limited data sample. Five models were built, with each model trained on 80% of the data and validated on the final 20%. Unlike the regular train-test split, this gives a less biased estimate of the model's skill on unseen data using a small dataset, and predictions can be made on all data. After calculating the standard deviation of their test accuracies, a low standard deviation implies that the algorithm and data are consistent and the model is more likely to generalize to the real world, and vice versa.

> **Classification algorithm:** All experiments used a batch size of 32 and a learning rate of 0.001.
> 1. Split the dataset into k groups
> 2. for fold i = 0, …, k **do**
>     a. Use 1 group as the validation set and the remaining groups as the training set
>     b. Fit a model on the training set for 200 epochs and evaluate on the validation set
>     c. Save the model with the lowest validation loss

Evaluation metrics: The five models were evaluated on the fixed test set and their average test accuracy was used as the main metric. The confusion matrices of all models were also calculated to compare the classifier performance for each individual class. To ensure that the classifiers were looking at regions with stool in the image, Gradient-weighted Class Activation Mapping (Grad-CAM) [6] was applied to the test images. Grad-CAM uses the gradients of any target concept flowing into the final convolutional layer to produce a localization heatmap highlighting important regions in the image to the classifier for class prediction. The average heatmap of each class was derived by averaging the heatmaps of all processed images in each class.

### 3.1 Two-Class Models vs Three-Class Models

Initially, our classifier models were trained on only two separate classes of images – normal stool and abnormal stool. However, after performing Grad-CAM, the average heatmaps produced highlighted that the models were not learning features of the stool.

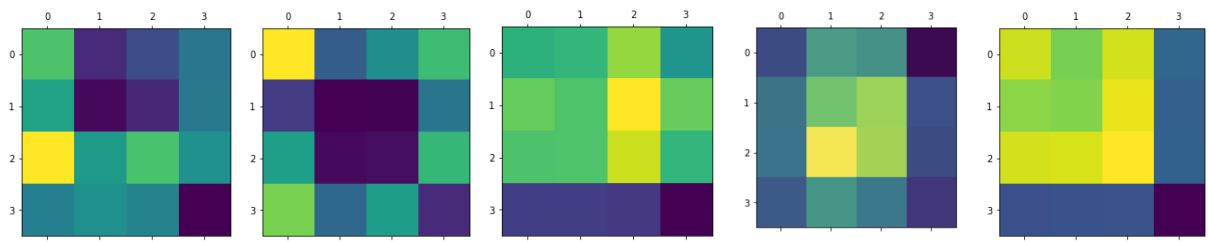

Fig 1: Average heatmaps of the Two-Class Models (Fig 1a Abnormal; Fig 1b Normal) in comparison to those of the Three-Class Models (Fig 1c Abnormal; Fig 1d Normal; Fig 1e No Stool). The areas with warm colours (yellow) signify high importance while those with cool colours (purple) signify low importance.

From the two-class average heatmaps above, yellow high-intensity areas are scattered at different edges of the image, but not at the centre. This means that the models were not learning the features of the stool itself, but the surroundings (toilet bowl, bathroom floor, etc.). To counter this, a third class - no stool (toilet bowls without stool) was introduced so that the model would place less focus on the background and instead focus on regions of the image with stool (centre).

## 3.2 Improving Classifier Performance with GANs

In order to choose the GAN which could generate the most realistic images of high fidelity to our original dataset, various GANs were trained to generate 128x128x3 images and their performance was evaluated with the Frechet Inception Distance (FID) Score metric [7, 8]. The FID Score is a popular GAN evaluation metric that calculates the Frechet Distance between the multivariate Gaussian distributions of the real and fake generated images. The multivariate Gaussian distributions are obtained from the activations of the images passed through the Inception V3 model.

$$FID = \| \mu_r - \mu_g \|^2 + Tr(\Sigma_r + \Sigma_g - 2\sqrt{\Sigma_r \Sigma_g})$$

Where $\mu_r$ and $\mu_g$ are the feature-wise mean of the real and generated images respectively, $\Sigma_r$ and $\Sigma_g$ are the covariance matrix for the real and generated feature vectors, and *Tr* is the trace linear algebra operation (sum of the elements along the main diagonal of the square matrix). The lower the FID Score, the lower the Frechet Distance between the real and generated images and the higher the fidelity of the generated images.

Below are the three GANs that were experimented (See Appendix for model architecture and training details):

Deep Convolutional (DCGAN) [9]: DCGANs are composed of convolution layers without max pooling or fully-connected layers. Convolutional strides are used for downsampling while transpose convolutions are used for upsampling. Three DCGANs were trained to generate images of each class.

Conditional GAN (CGAN) [10]: A limitation of GANs is its inability to control the class of images generated. CGANs overcome this by enabling conditional image generation, by adding the image's class label to the model. Learning the features of each class allows the generation of images more similar to the dataset distribution. A CGAN was trained on all three classes.

Differentiable Augment (DiffAugment) StyleGAN2 [11]: Aimed to combat limited GAN training data, DiffAugment applies differentiable augmentation techniques (translation, cut-out augmentation and colour jittering) to both real and generated images before passing it to the GAN discriminator. This has enabled GANs trained on only 100 images to generate realistic images without suffering mode collapse, which is suitable for our insufficient dataset. Only differentiable augmentation techniques are used to prevent the generator from matching the distribution of the augmented images and generating distorted images. We trained three identical DiffAugment StyleGAN2 for each class.

The final decision was to use the DiffAugment StyleGAN2 as it generated the most realistic images and had the lowest FID Score (see Section 4). To investigate if the addition of GAN-generated images can indeed improve the accuracy of classifiers, two types of models were trained – models trained only on the dataset, and models trained on both the dataset and 200 synthetic images generated on-the-fly every fold for each class by the DiffAugment StyleGAN2.

### 3.3 Further Improvements with Image Augmentation

To further boost classifier performance, image augmentation [12] was applied to all training images (random rotation of 10°, height and width shifts of range 0.1, horizontal flips).

## 4   Results and Discussion

### 4.1 GAN Results

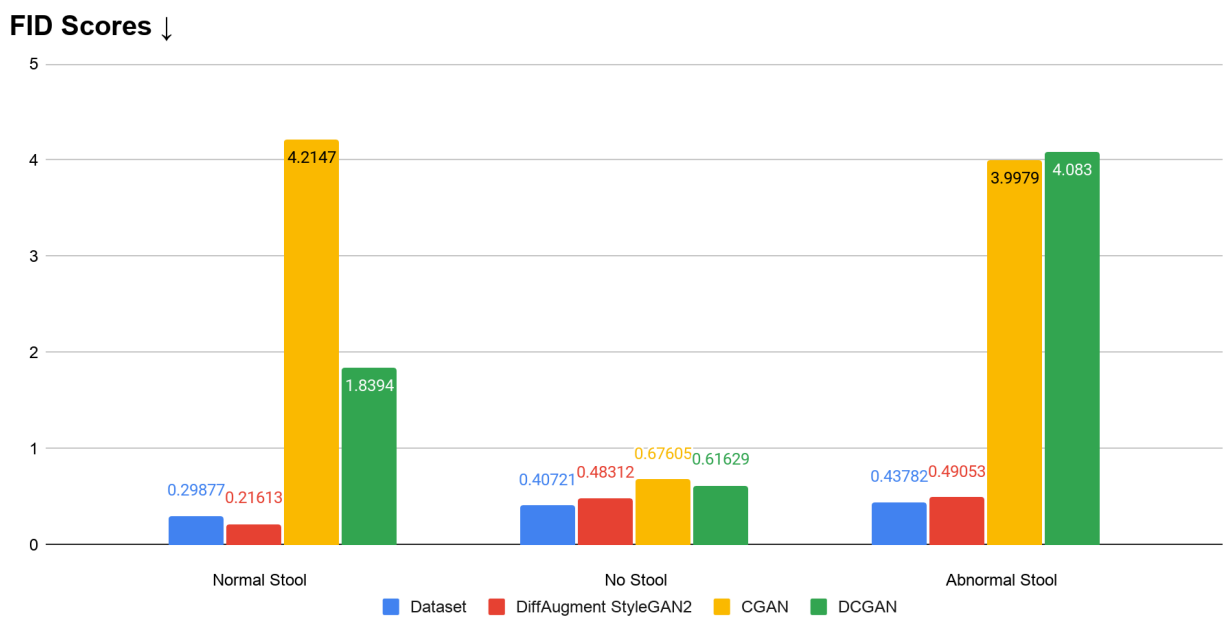

Fig 2: Graph showing the FID scores for the various GANs for each class of images.

The DiffAugment StyleGAN2 (red) achieved the lowest FID scores for all classes, implying that the synthetic images generated by this GAN have the highest fidelity to the dataset. This conclusion was further corroborated with a visual assessment of the GAN-generated images, where synthetic images generated by the DiffAugment StyleGAN2 were the most realistic and did not suffer from mode collapse (See appendix). This proves that differentiable augmentation can enable GANs to generate realistic images of a wide variety even with limited data.

**4.2 Classification Results**

| Model Type | GAN Images | Mean Test Accuracy / % | Best Accuracy Model / % | Mean Loss |
|---|---|---|---|---|
| Two Classes | x | 75.33 ± 3.27 | 80.00 | 0.64 |
|  | ✓ | 78.17 ± 2.64 | 82.50 | 0.54 |
| Three Classes | x | 79.88 ± 3.48 | 85.62 | 0.55 |
|  | ✓ | 84.25 ± 2.07 | 86.87 | 0.48 |
| Three Classes + Image Augmentation | x | 87.13 ± 2.55 | 90.62 | 0.38 |
|  | ✓ | 90.38 ± 2.04 | 94.38 | 0.30 |

Fig 3: Table showing the performance of some of the models. See appendix for additional results.

From the table above, classifier models trained with additional synthetic images generated by the DiffAugment StyleGAN2 achieved higher mean test accuracies compared to those trained only on our dataset by around 3-5%. This proves our hypothesis that the use of GANs can improve the performance of classifier models by increasing the size of the dataset. We believe that the classifier performance will improve when the number of additional synthetic images increases.

The model was later deployed using TensorflowLite to an Android mobile app - Poolice. It was created with an easy-to-use interface and a built-in camera function, achieving our original goal of assisting potential colorectal cancer patients in assessing if they have signs of symptoms.

## 5 Implications and Conclusion

In this research, we discovered that the use of GANs can improve the performance of a classifier by increasing the size of the dataset. With the following discovery, our stool recognition neural network achieved a high test accuracy of 94% (Three-Class Image Augmentation with GAN images). Moving on, we intend to study if classifier models can achieve comparable accuracies with unsupervised learning techniques that do not require any training data.

## 6 Acknowledgements

We would like to thank our teacher, Mr Shaun De Souza and mentors, Dr Shen Bingquan and Ms Wong Minn Xuan from DSO National Laboratories for their invaluable guidance and advice through this research project, as well as our family members and friends for contributing pictures of their stool for our dataset.

# 8 Appendix

## 8.1 Samples of Dataset Images

Normal Stool

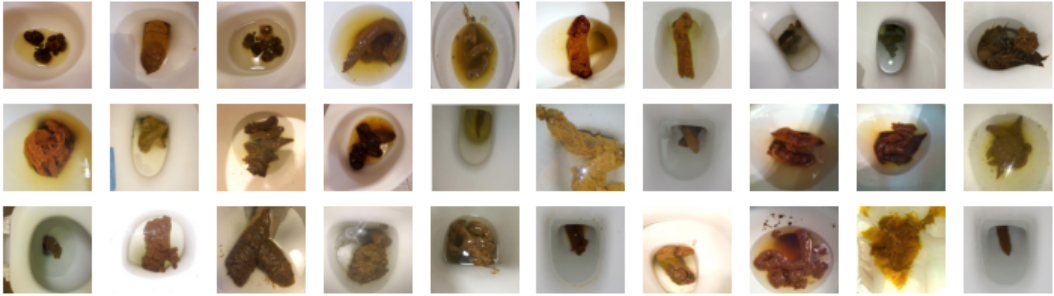

Fig 4a: Images from the Normal Stool class

Abnormal Stool

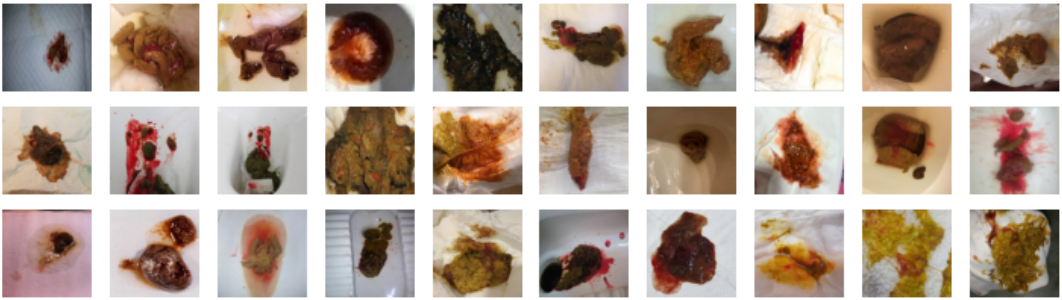

Fig 4b: Images from the Abnormal Stool class

No Stool

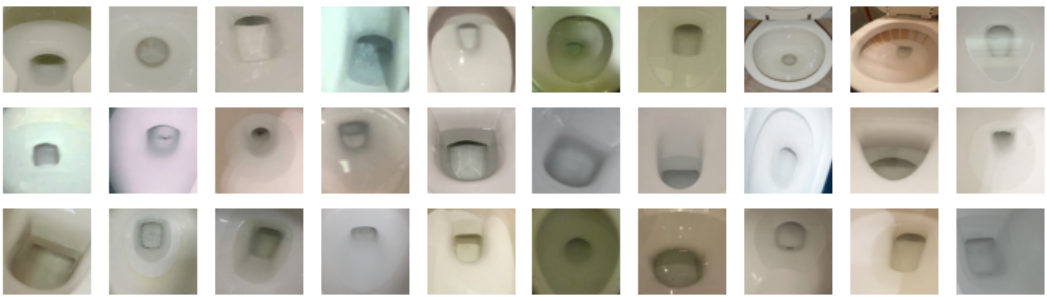

Fig 4c: Images from the No Stool class

## 8.2 Samples of GAN-Generated Images

Normal Stool

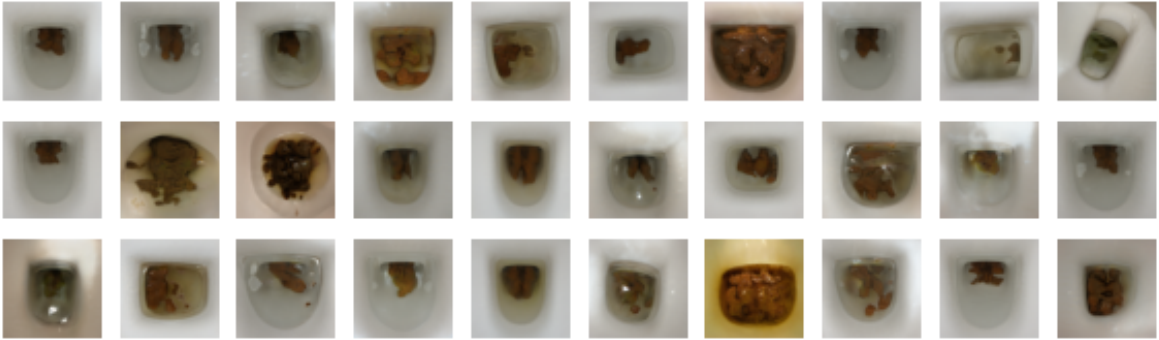

Fig 5a: Normal stool images generated by DiffAugment StyleGAN (Most Realistic)

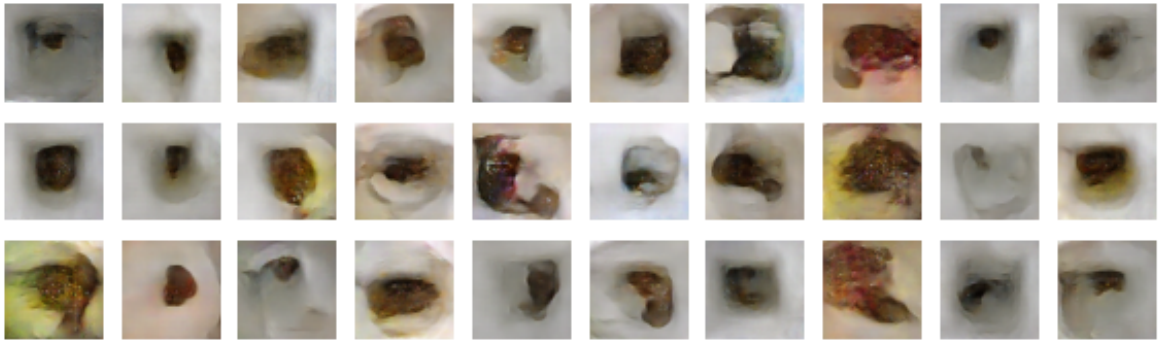

Fig 5b: Normal stool images generated by CGAN after 100 epochs

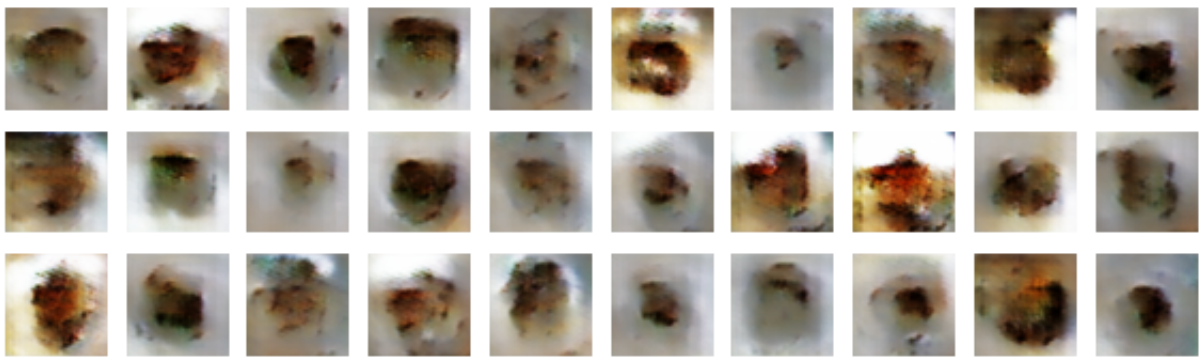

Fig 5c: Normal stool images generated by DCGAN after 100 epochs

Abnormal Stool

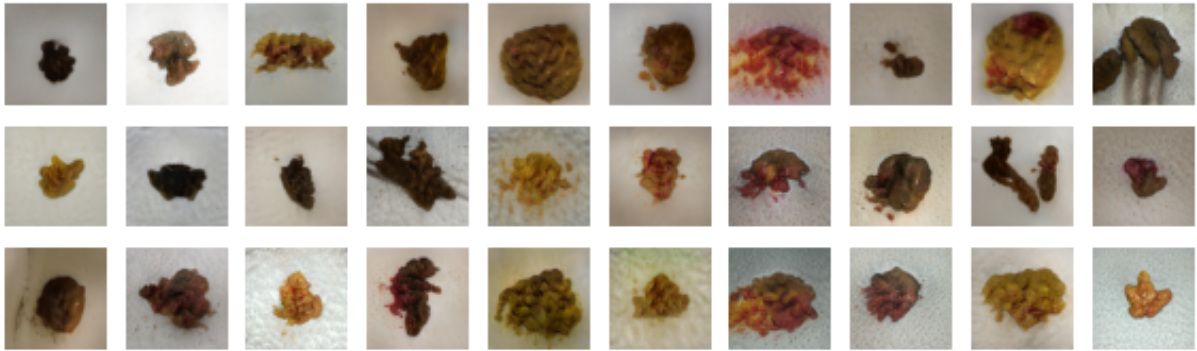

Fig 6a: Abnormal stool images generated by DiffAugment StyleGAN (Most Realistic)

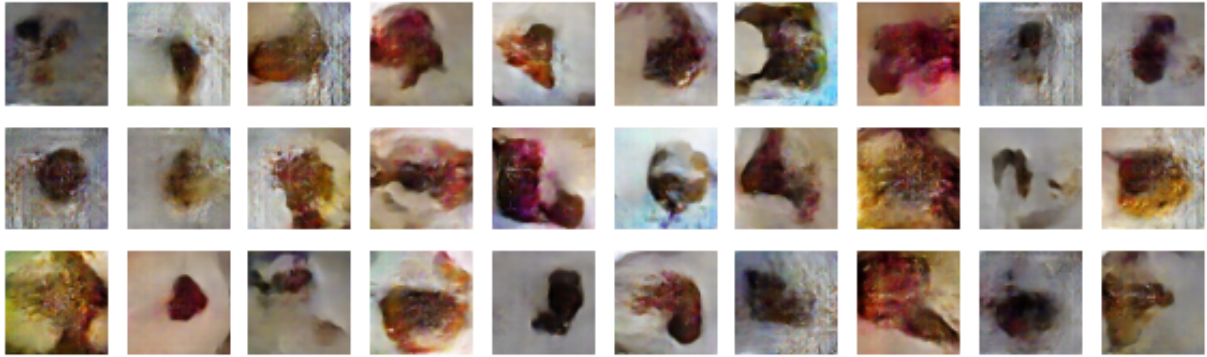

Fig 6b: Abnormal stool images generated by CGAN after 100 epochs

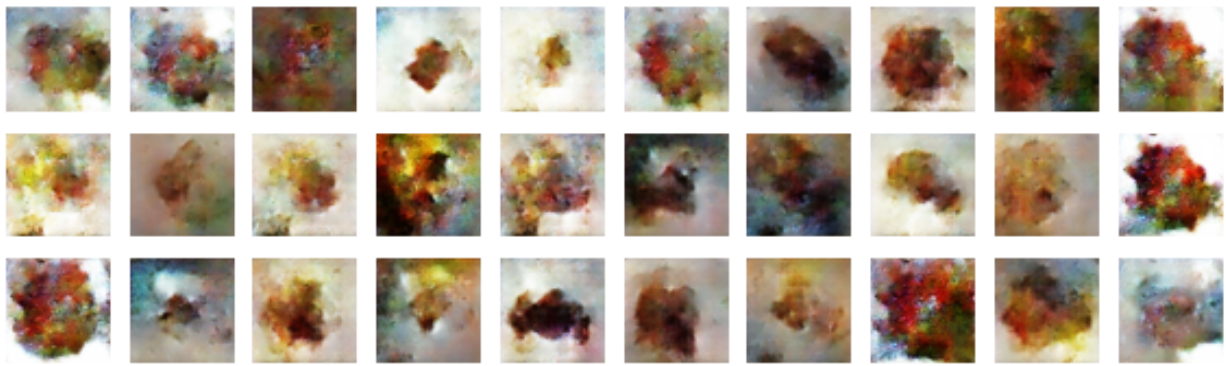

Fig 6c: Abnormal stool images generated by DCGAN after 100 epochs

No Stool

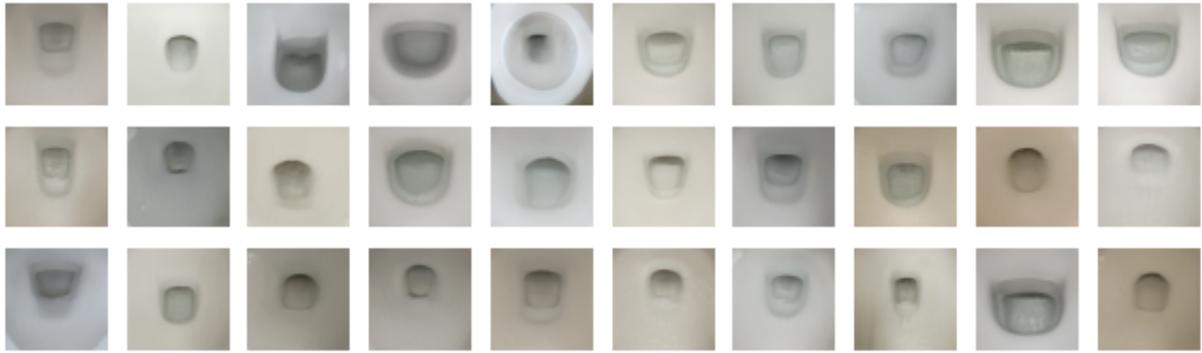

Fig 7a: No stool images generated by DiffAugment StyleGAN (Most Realistic)

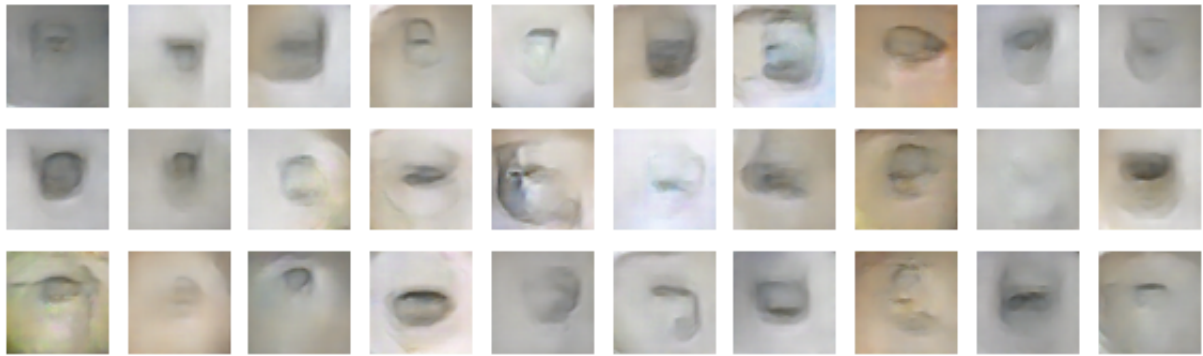

Fig 7b: No stool images generated by CGAN after 100 epochs

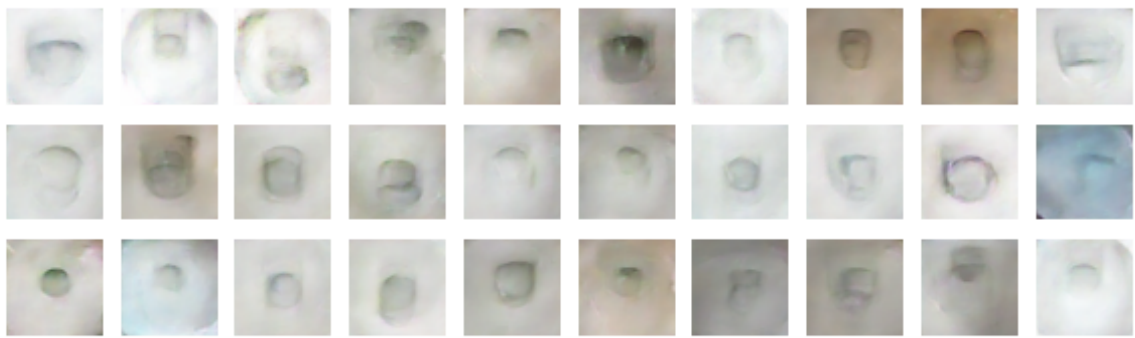

Fig 7c: No stool images generated by DCGAN after 100 epochs

## 8.3 Classification Results

Model 1: Similar dataset images
Model 2: Similar dataset images + GAN images
Model 3 (in report): All dataset images
Model 4 (in report): All dataset images + GAN images

| Model Type | Model | Mean Test Accuracy / % | Best Accuracy Model / % | Mean Loss |
|---|---|---|---|---|
| Two Classes | 1 | 63.83 ± 0.85 | 65.00 | 10.78 |
|  | 2 | 71.67 ± 4.31 | 76.67 | 4.20 |
|  | 3 | 75.33 ± 3.27 | 80.00 | 0.64 |
|  | 4 | 78.17 ± 2.64 | 82.50 | 0.54 |
| Three Classes | 1 | 72.88 ± 1.51 | 75.63 | 8.81 |
|  | 2 | 79.38 ± 3.09 | 82.50 | 2.96 |
|  | 3 | 79.88 ± 3.48 | 85.62 | 0.55 |
|  | 4 | 84.25 ± 2.07 | 86.87 | 0.48 |
| Three Classes + Image Augmentation | 1 | 72.38 ± 0.73 | 73.75 | 6.14 |
|  | 2 | 78.75 ± 0.64 | 82.50 | 2.05 |
|  | 3 | 87.13 ± 2.55 | 90.62 | 0.38 |
|  | 4 | 90.38 ± 2.04 | 94.38 | 0.30 |

Fig 8: Table showing the performance of all classification models

## 8.4 Confusion Matrices

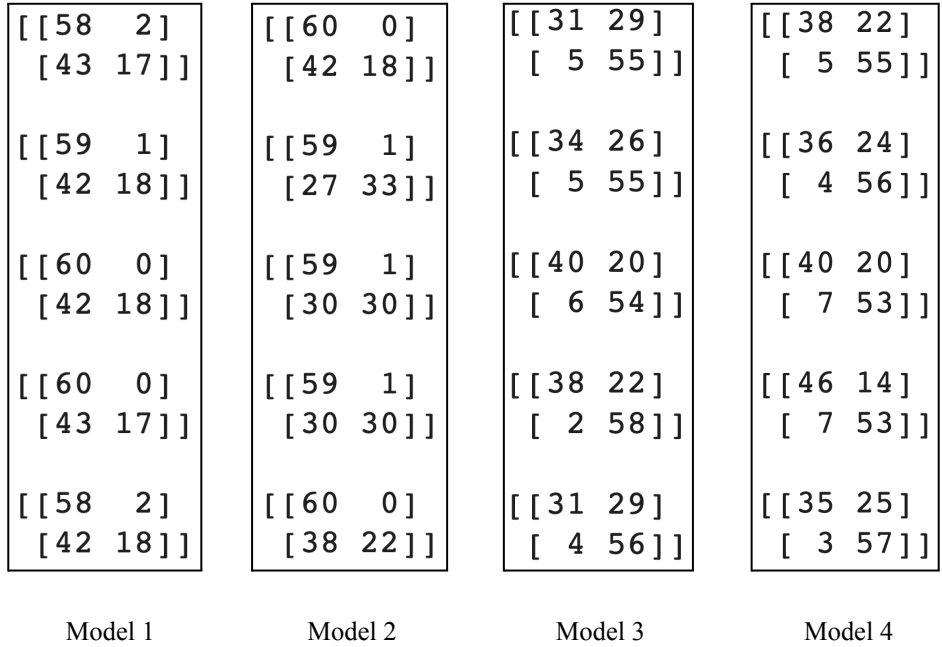

Fig 9: Confusion Matrices of Two-Class Models (Abnormal: Row 1 Col 1, Normal: Row 2 Col 2)

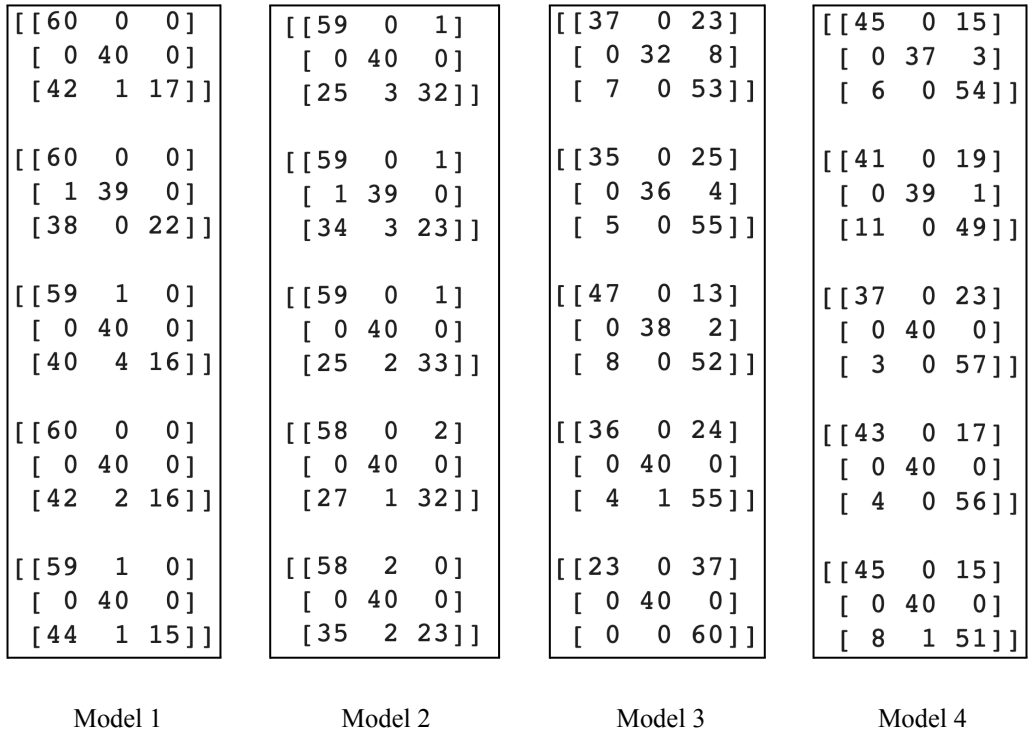

Fig 10: Confusion Matrices of Three-Class Models
(Abnormal: Row 1 Col 1, No stool: Row 2 Col 2, Normal: Row 3 Col 3)

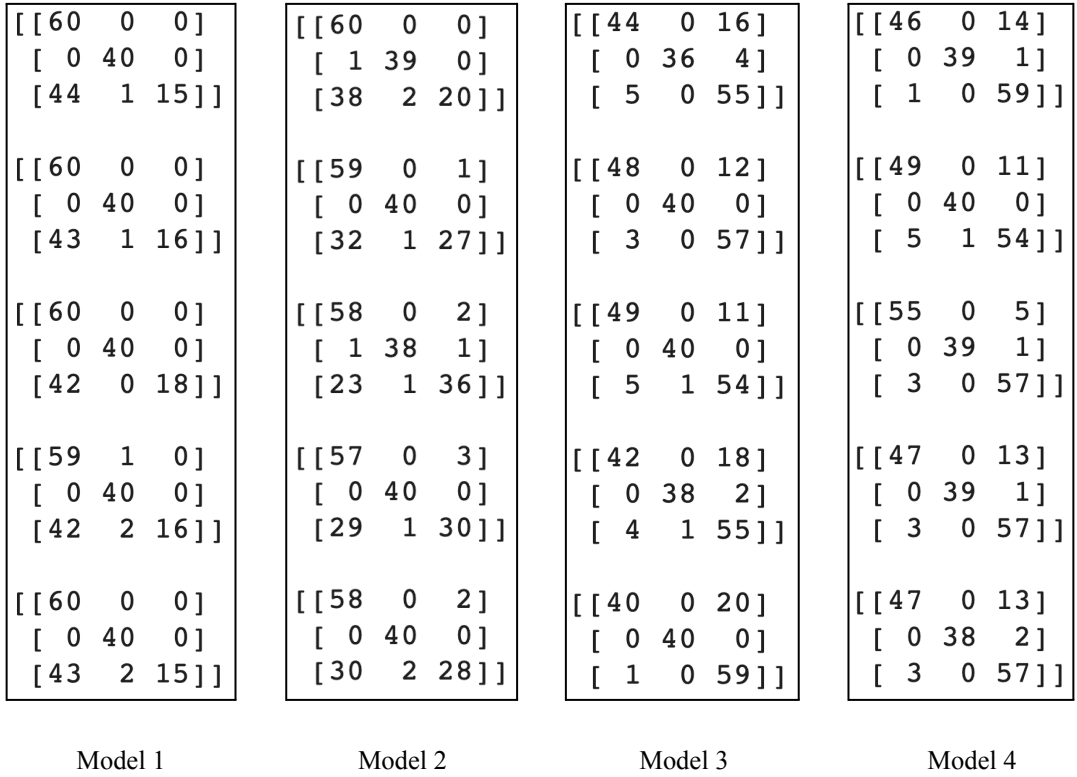

```
[[60  0  0]      [[60  0  0]      [[44  0 16]      [[46  0 14]
 [ 0 40  0]       [ 1 39  0]       [ 0 36  4]       [ 0 39  1]
 [44  1 15]]      [38  2 20]]      [ 5  0 55]]      [ 1  0 59]]

[[60  0  0]      [[59  0  1]      [[48  0 12]      [[49  0 11]
 [ 0 40  0]       [ 0 40  0]       [ 0 40  0]       [ 0 40  0]
 [43  1 16]]      [32  1 27]]      [ 3  0 57]]      [ 5  1 54]]

[[60  0  0]      [[58  0  2]      [[49  0 11]      [[55  0  5]
 [ 0 40  0]       [ 1 38  1]       [ 0 40  0]       [ 0 39  1]
 [42  0 18]]      [23  1 36]]      [ 5  1 54]]      [ 3  0 57]]

[[59  1  0]      [[57  0  3]      [[42  0 18]      [[47  0 13]
 [ 0 40  0]       [ 0 40  0]       [ 0 38  2]       [ 0 39  1]
 [42  2 16]]      [29  1 30]]      [ 4  1 55]]      [ 3  0 57]]

[[60  0  0]      [[58  0  2]      [[40  0 20]      [[47  0 13]
 [ 0 40  0]       [ 0 40  0]       [ 0 40  0]       [ 0 38  2]
 [43  2 15]]      [30  2 28]]      [ 1  0 59]]      [ 3  0 57]]
```

     Model 1              Model 2             Model 3             Model 4

Fig 11: Confusion Matrices of Three-Classes + Image Augmentation Models
(Same rows and columns as 3-Class Models)

## 8.5 Average Grad-CAM Heatmaps

Two-Class Models

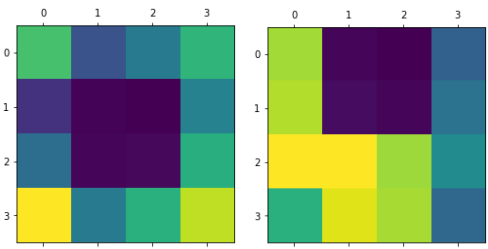
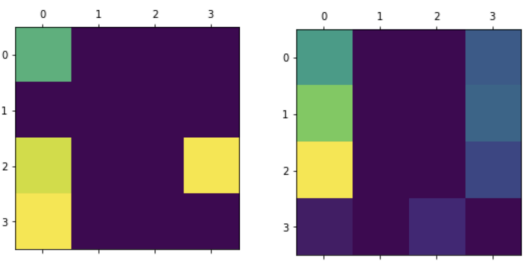

Fig 12: Average heatmaps of Two-Class Model 1      Fig 13: Average heatmaps of Two-Class Model 2
(Fig 12a Abnormal; Fig 12b Normal)              (Fig 13a Abnormal; Fig 13b Normal)

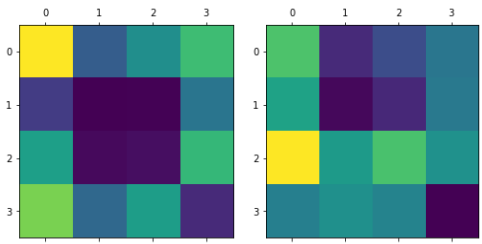

Fig 14: Average heatmaps of Two-Class Model 3
(Fig 14a Abnormal; Fig 14b Normal)

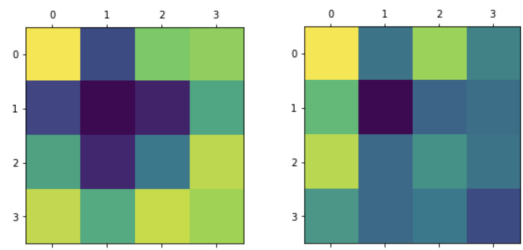

Fig 15: Average heatmaps of Two-Class Model 4
(Fig 15a Abnormal; Fig 15b Normal)

3-Class Models

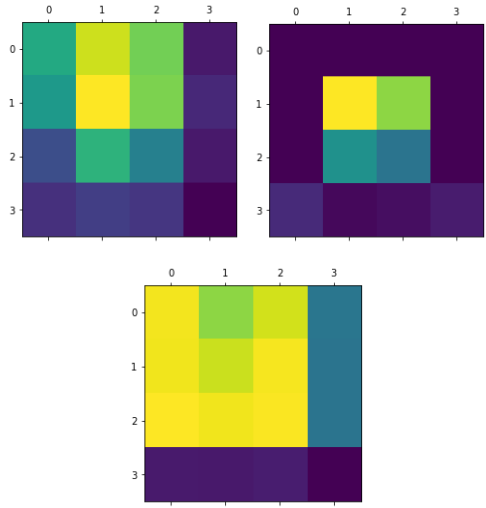

Fig 16: Average heatmaps of Three-Class Model 1
(Fig 16a Abnormal; Fig 16b Normal; Fig 16c No stool)

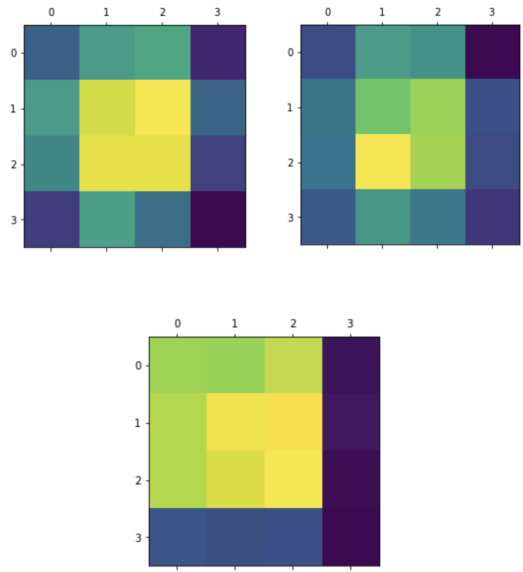

Fig 17: Average heatmaps of Three-Class Model 2
(Fig 17a Abnormal; Fig 17b Normal; Fig 17c No Stool)

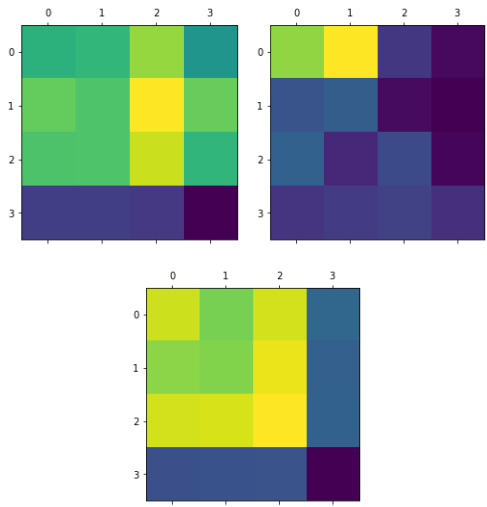

Fig 18: Average heatmaps of Three-Class Model 3
(Fig 18a Abnormal; Fig 18b Normal; Fig 18c No stool)

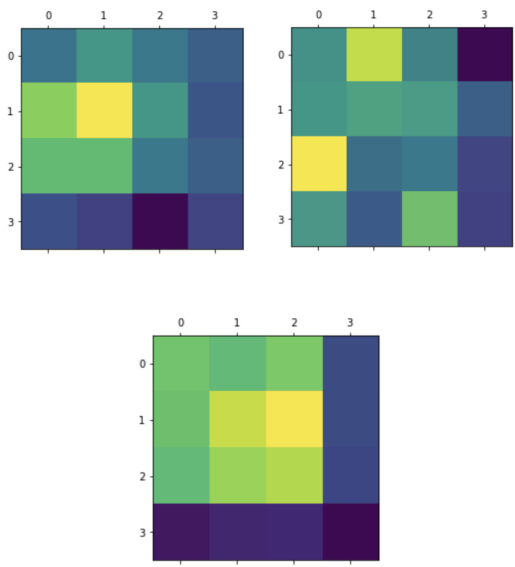

Fig 19: Average heatmaps of Three-Class Model 4
(Fig 19a Abnormal; Fig 19b Normal; Fig 19c No Stool)

3-Class Models with Image Augmentation

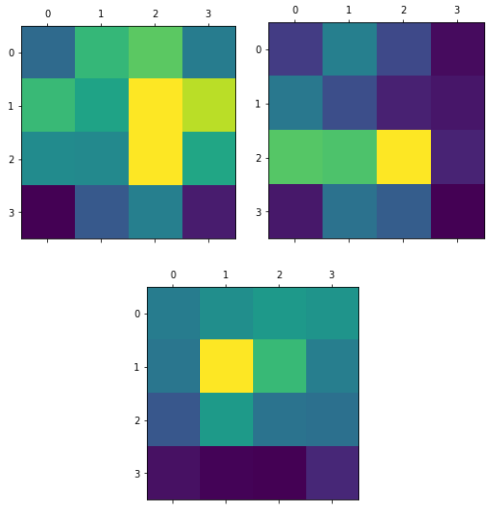

Fig 20: Average heatmaps of Three-Class Model 1
with Image Augmentation
(Fig 20a Abnormal; Fig 20b Normal; Fig 20c No stool)

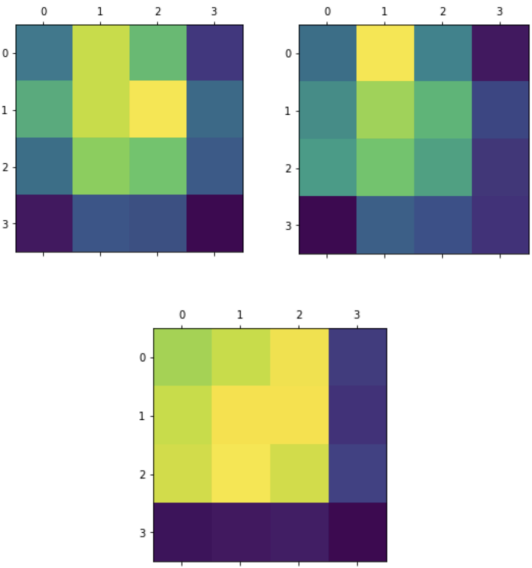

Fig 21: Average heatmaps of Three-Class Model 2
with Image Augmentation
(Fig 21a Abnormal; Fig 21b Normal; Fig 21c No stool)

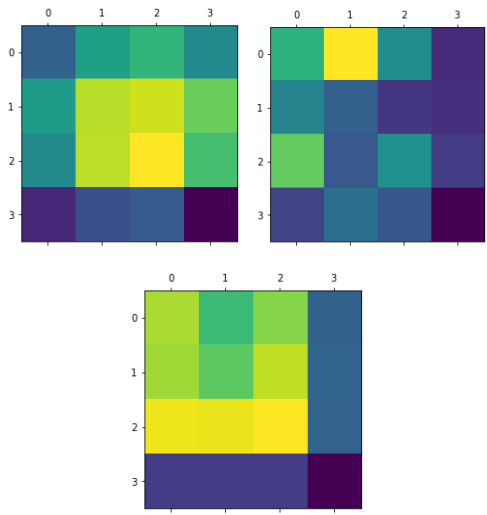

Fig 22: Average heatmaps of Three-Class Model 3
with Image Augmentation
(Fig 22a Abnormal; Fig 22b Normal; Fig 22c No stool)

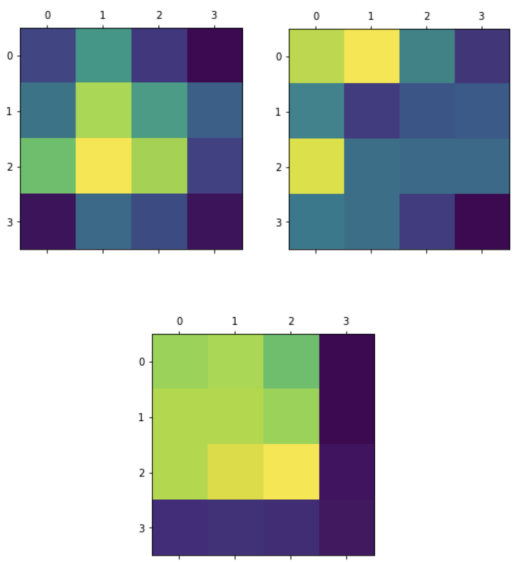

Fig 23: Average heatmaps of Three-Class Model 4
with Image Augmentation
(Fig 23a Abnormal; Fig 23b Normal; Fig 23c No stool)

## 8.6 Model Architectures

GANs

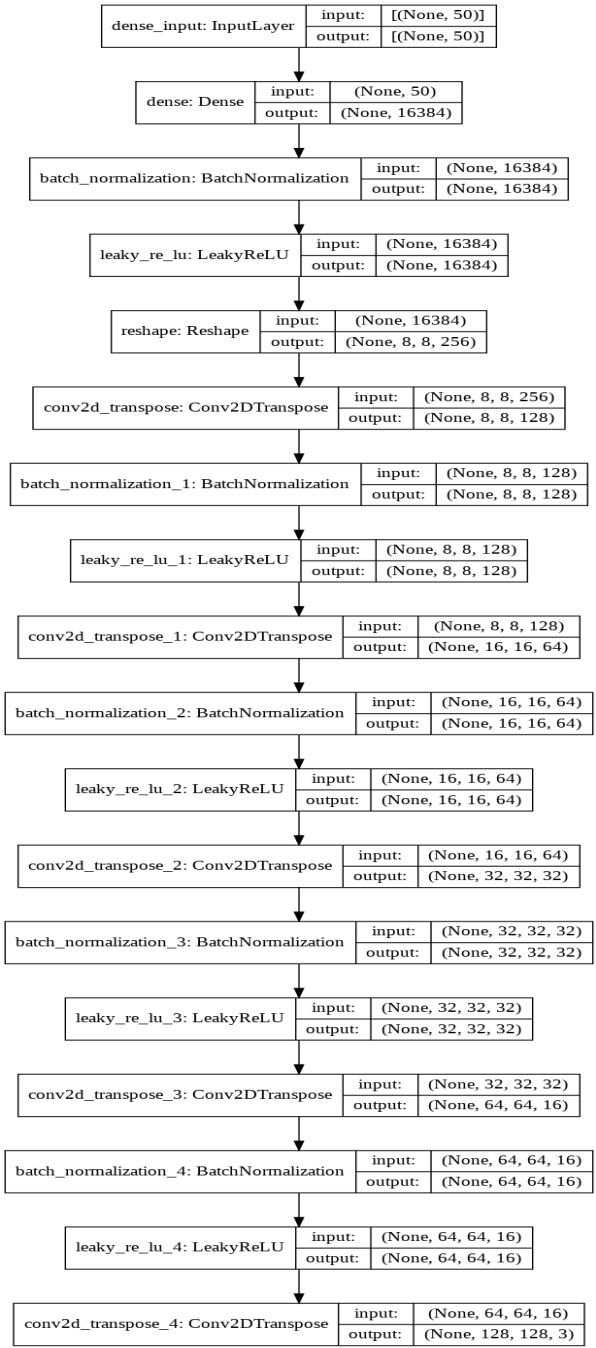

Fig 24: Model Architecture of the DCGAN

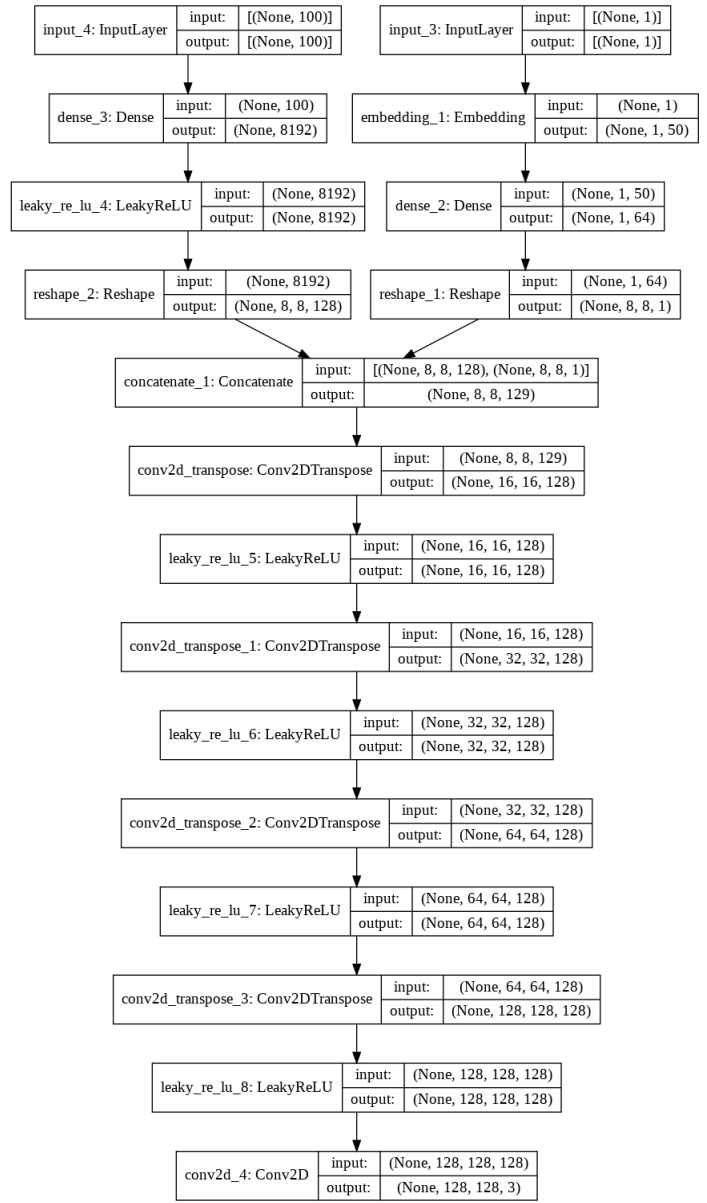

Fig 25: Model Architecture of the CGAN

Classifiers

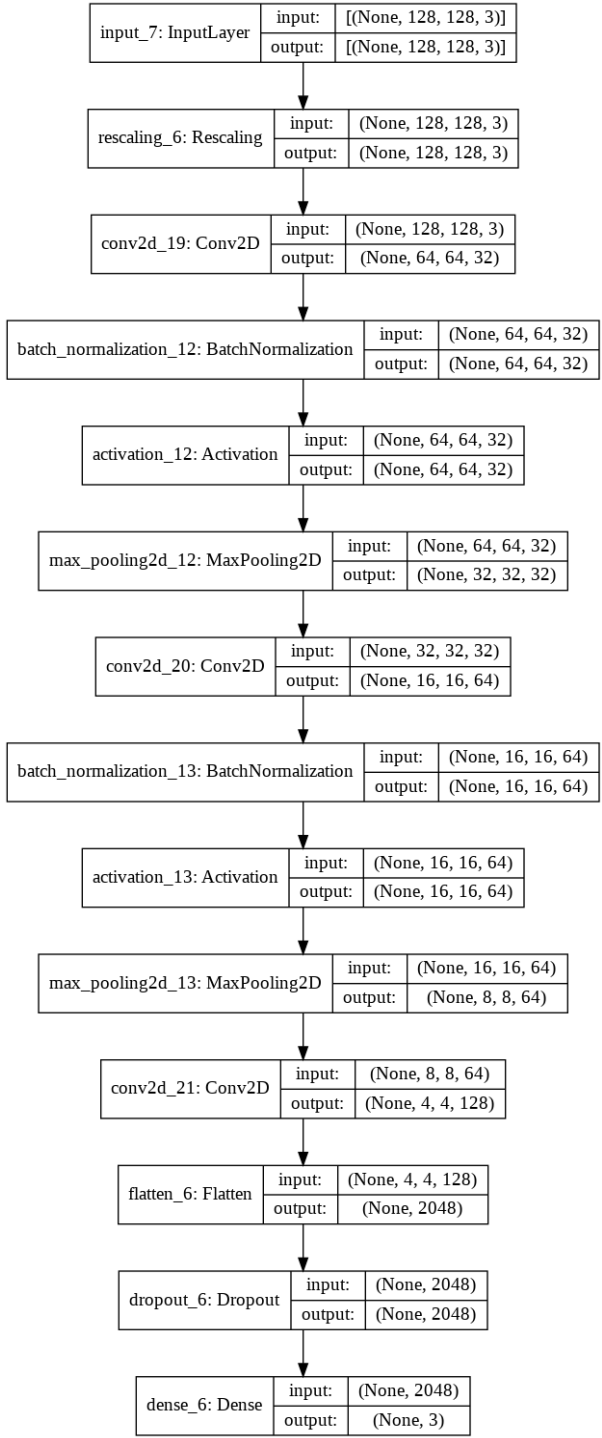

Fig 26: Model Architecture of the Two-Class Model

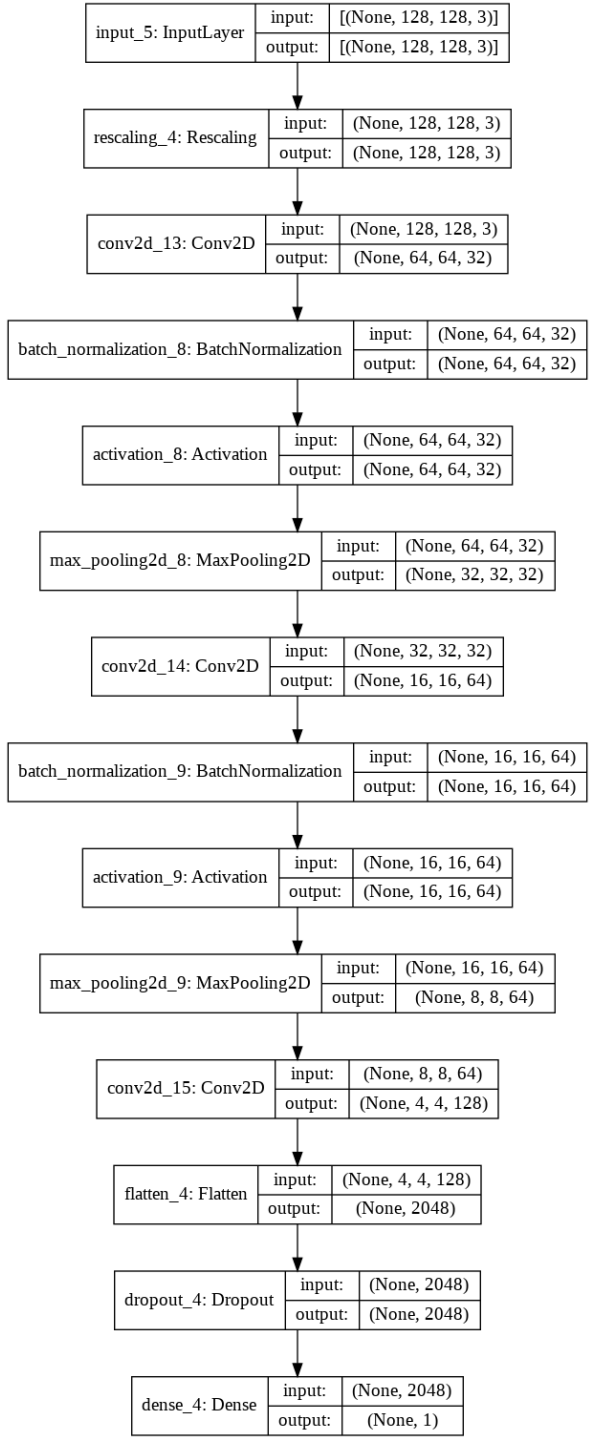

Fig 27: Model Architecture of the Three-Class Model

## 8.7 Screenshots of Poolice Android App

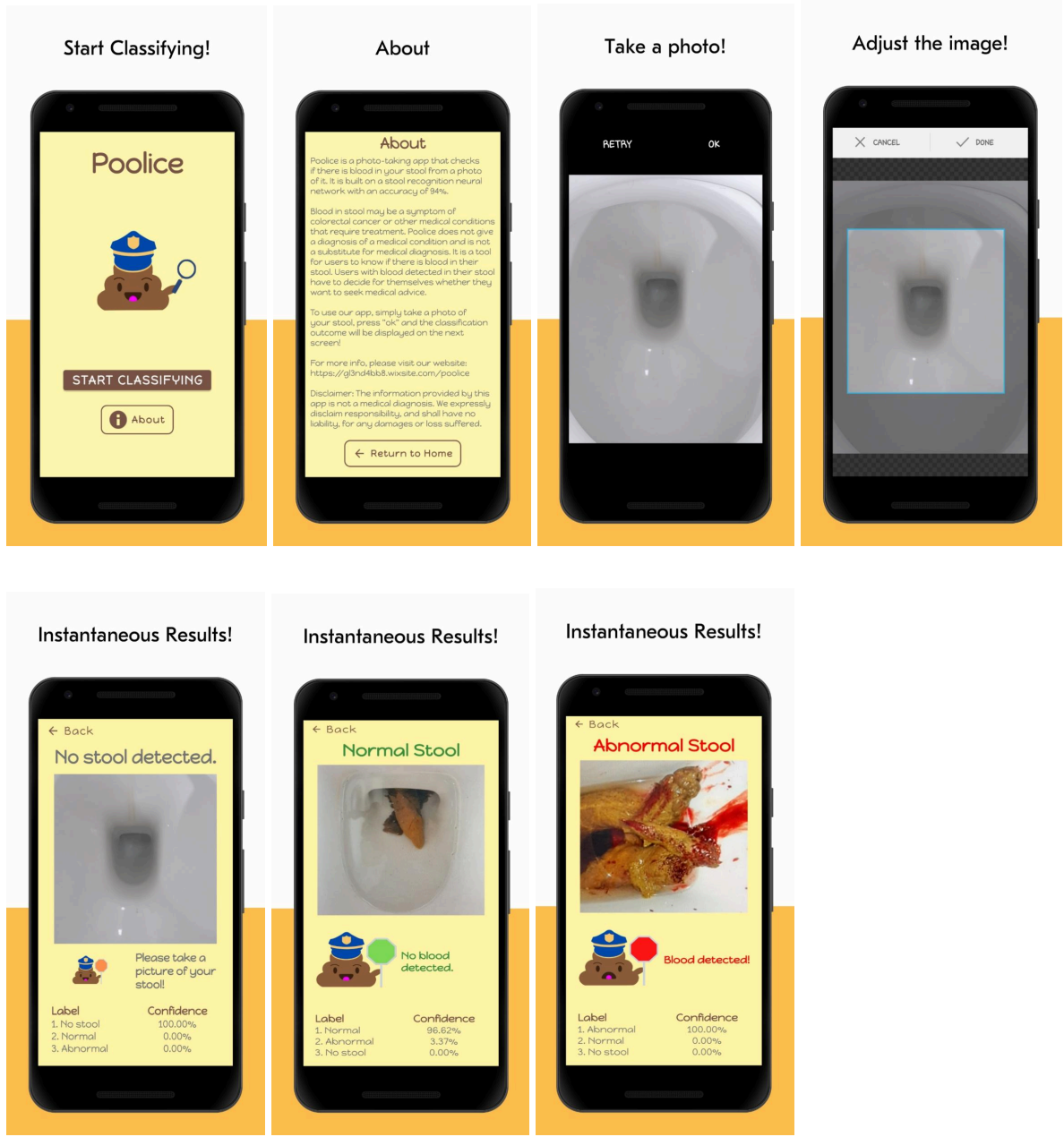

Fig 28 (a) to (g): Screenshots of the Android mobile app, Poolice, showing the interfaces the user will be viewing, including the diagnosis at the end